\documentclass[aps,cyrwin,prl,twocolumn,floats,graphicx,epsfig,euscript,amsmath,amssymb]{revtex4}
\usepackage[dvips]{graphicx}
\usepackage{epsfig}
\DeclareMathAlphabet\EuScript{U}{eus}{m}{n} \SetMathAlphabet\EuScript{bold}{U}{eus}{b}{n}

\def\lapprox{\,\raise0.4ex\hbox{$<$}\kern-0.8em\lower0.7ex\hbox{$\sim$}\,}
\def\gapprox{\,\raise0.4ex\hbox{$>$}\kern-0.8em\lower0.7ex\hbox{$\sim$}\,}
\begin{document}
\title
{Goldstone mode stochastization in quantum Hall ferromagnet}

\author{$\qquad$ A.V. Larionov, L.V. Kulik, S. Dickmann, and I.V. Kukushkin}

\affiliation{$$Institute of Solid State Physics, Russian
Academy of Sciences, Chernogolovka, 142432 Russia}

\date{\today}

\begin{abstract}
Experimental and theoretical studies of the coherent spin dynamics of two-dimensional GaAs/AlGaAs electron gas were performed.  The system in the quantum Hall ferromagnet state exhibits a spin relaxation mechanism that is determined by many-particle Coulomb interactions. In addition to the spin exciton with changes in the spin quantum numbers of $\delta S\!=\!\delta S_z
\!=\!-1$, the quantum Hall ferromagnet supports a Goldstone spin exciton that changes the spin quantum numbers to $\delta S\!=\!0$ and $\delta S_z\!=\!-1$, which corresponds to a coherent spin rotation of the entire electron system to a certain angle. The Goldstone spin exciton decays through a specific relaxation mechanism that is unlike any other collective spin state.
\vskip 1mm

\end{abstract}
\maketitle
\vspace{-20mm}{\em Introduction.} Spin relaxation mechanisms in two-dimensional (2D) electron systems have not yet been elucidated due to the large number of competing mechanisms and the
complex effects of the many-particle Coulomb interactions on relaxation. 2D confinement and the quantizing magnetic field ensure a cardinal
rearrangement of the electron energy spectrum, effectively making it zero-dimensional.
Standard single-particle relaxation channels (see, e.g., Ref. \cite{Fabian} and the references therein) are suppressed, which prolongs spin relaxation time. On the other hand, electron-electron correlations, very essential in the case, make the spectrum again two-dimensional. At integer filling factors and at some fractional ones the simplest electron excitations are magnetoexcitons \cite{go68} with well defined 2D momenta, specifically representing magnetoplasmons, spin waves,
or spin-cyclotron excitons \cite{Bychkov81,pi92,ga08,va06,dr10,wurst}. New spin relaxation mechanisms, e.g., related to the exciton-exciton scattering processes appear.

The most comprehensive concept of spin relaxation was developed for the quantum Hall ferromagnet (QHF), $\nu\!=\!2n\!+\!1$ \cite{di04,di96,Fukuoka,Zhuravlev14,di12}, in which the $n\!-\!1$ low Landau levels at $T\!\to\!0$ are fully occupied and the $n$th level is filled by spin-up electrons aligned along ${\vec B}$. The QHF is in fact a high symmetry system for investigating the influence of many-particle Coulomb interactions on the spin excitation spectrum \cite{Bychkov81,pi92,ga08,va06,dr10}. Research into the nonequilibrium spin system of the QHF is also a direct method of evaluating the influence of the many-particle Coulomb interactions on spin relaxation in 2D systems.

Deviation of the spin system from equilibrium for the QHF can be described as formation of spin excitons comprising an effective hole in a spin-polarized electron
system and an electron with an opposite spin \cite{Bychkov81}.  Formation of a zero momentum spin
exciton (Goldstone spin exciton) changes the spin projection along the
magnetic field $S_z$ but maintains the total spin of the electron system $S$. Thus the presence of Goldstone excitons corresponds to a coherent spin rotation
about the ${\vec B}$ direction. {The stochastization and simultaneously the relaxation to the ground state of such a Goldstone mode (stationary eigen state), both governed by one type of relaxation mechanisms, were theoretically considered earlier \cite{di04,di96}. In particular, in Ref. \cite{di04} the relaxation was supposed to occur via the mechanism of spin-orbit coupling affected by the smooth random potential that always takes place in 2D systems. (In fact, a similar relaxation channel is realized for nonzero momentum spin excitons where $S$ and $S_z$ are equally reduced and where the relaxation was studied not only theoretically \cite{di12} but also experimentally \cite{Zhuravlev14}.)}

\vspace{-.2mm}

Here we consider a different situation. Initially using optical excitation we create a non-stationary state where the entire electron spin is rotated as a whole about its equilibrium direction. The coherent Goldstone mode arises if the following condition occurs: $|\delta S|\!<\!|\delta S_z|$. This state can be mathematically described
by the action of the operator \vspace{-.5mm}
${\hat S}_-\!=\!{\hat S}_x\!-\!i{\hat S}_y$ onto \vspace{-.0mm}
 the ground state
$|{\rm 0,0}\rangle\!=\!|\overbrace{\uparrow\uparrow\uparrow...\uparrow}^{\hphantom{N}
{}_{{}_{{}_{{\cal N}_\phi}}}}\,\rangle$
(${\cal N}_\phi$ is the degeneration number of the completely occupied Landau level).
The $N$-fold action of this operator: $|N,0\rangle\!=\!({\cal Q}^\dag_0)^N|0,0\rangle$
(we use the designation ${\cal Q}^\dag_0\!=\!S_-\!/\!\sqrt{{\cal N}_\phi}$) represents
an eigenstate, the Goldstone condensate, i.e., the state where
$S\!=\!{\cal N}_\phi/2$ and $S_z\!=\!{\cal N}_\phi/2\!-\!N$. All the spins in this
condensate are tilted as a whole from the ${\vec B}$ direction by angle $\theta$: $\cos\theta\!=\!S_x/S$.
The energy of the state calculated from the ground state level is equal
to $E_N\!=\epsilon_ZN$ ($\epsilon_Z\!=\!|g|\mu_BB$ is the electron Zeeman energy), so the total
form of the basis Goldstone state is $e^{-iE_Nt}|N,0\rangle$.\vspace{-2mm}

An elementary stochastization process represents a change from state $|N,0\rangle$ to a state where one of the zero excitons becomes nonzero:
$|N,{\bf q}\rangle=$ ${}\!({\cal Q}^\dag_0)^{N-1}{\cal Q}_{\bf q}^\dag|0,0\rangle$, where
  ${\cal Q}_{\bf q}^{\dag}=$ ${\cal N}_{\phi}^{-1/2}\sum_{p}
  e^{-iq_x p}
  b_{p+\frac{q_y}{2}}^{\dag}a_{p-\frac{q_y}{2}}\,$. $a_p$ and $b_p$ are the Fermi annihilation operators
corresponding to the electron states on the upper Landau level with spin-up
($a\!=\uparrow$) and spin-down ($b\!=\downarrow$).
Both $|N,0\rangle$ and $|N,{\bf q}\rangle$ are the QHF {\em eigenstates}. They
are orthogonal due to the translation invariance. The $|N,0\rangle\!\to\!|N,{\bf q}\rangle$ transition occurs without a change in the $S_z\!=\!{\cal N}_\phi/2\!-\!N$ component.
At $q\to 0$, the energies of both states ($E_N$) are also the same. However, the
states $|N,0\rangle$ and $|N,{\bf q}\rangle$ remain different even at $q\to 0$ (see discussion in Ref. \cite{di04}) and have different spin numbers: $S={\cal N}_\phi/2$ and $S={\cal N}_\phi/2\!-\!1$, respectively. Therefore the spin
tilt angle $\theta$ diminishes with this transition.

An approach describing the relaxation of Goldstone spin-excitons via two stages is developed. A fast stochastization stage (with characteristic time $\sim\!1\!-\!10\,$ns) converts Goldstone `zero' excitons into `nonzero' spin-excitons with the same energy; and the total number of spin excitons during this process is kept constant. The second stage is earlier  established \cite{Zhuravlev14} long-time relaxation ($\sim 100\,$ns) to the ground state, i.e. the `nonzero' spin-excitons annihilation governed by the spin-orbit coupling and smooth electrostatic random potential. Yet, the fast stochastization (dephasing of the spin precession) is not related to any previously identified relaxation mechanisms. The coherent spin precession decays because individual electron spins precess with slightly different Larmor frequencies in a spatially non-uniform environment. The spin-orbit coupling is irrelevant since it cannot provide transition from the zero excitons to the nonzero ones conditioned by the exciton total number conservation. We suppose that the spin-component $S$ nonconservation and the irreversibility of the process are provided by the $g$-factor long-wave spatial fluctuations.

{\em Experiment.}
To minimize the influence of random potential and separate the influence
of interparticle Coulomb interactions on spin relaxation, we studied high-quality
GaAs/AlGaAs heterostructures with single quantum wells (QWs) containing highly mobile
2D electron
gas ($\mu_{e}\!\simeq\!10^{7}\,$cm$^{2}$/Vs) with `dark' concentration $n_{s}\!\simeq \!0.7\!\times\!10^{11}\,$cm$^{-2}$ (sample A) and  $n_{s}\!\simeq\!2.4\!\times\!10^{11}\,$cm$^{-2}$ (sample B with $\mu_{e} \simeq 4\!\times\!10^{6}\,$cm$^{2}$/Vs). The spin dynamics was studied using the Kerr rotation technique at a base temperature of $1.5\,$K. The photoexcitation source was a picosecond titanium-sapphire laser with a tunable spectral width, and the wavelength of the pump laser beam coincided with that of the probe beam. The mean pump power was $\simeq\!1\,$mW, the laser spot size being on the order of $1\,$mm (the number
of pumped electrons did not exceed $\!10^{10}\,$cm$^{-2}$).
The samples were placed into an optical cryostat with a split solenoid
at 45 degrees in reference to the ${\vec B}$ direction, while the excited electron
spins were oriented close to normal to the sample surface due to the difference
between the refractive indices of GaAs and helium (see insert for Fig.1). The experimental geometry reproduced basically the arrangement first used in Ref. \cite{Fukuoka}
except that our setup enabled us to excite the electrons with a high spectral
resolution ($0.7\,$meV). It is extremely important not to mix up spin dynamics from different energy states if trying to separate many-particle and single-particle spin relaxation mechanisms.
The refraction geometry ensured that the generated spin excitations had a zero transverse (along the QW plane) momentum, i.e., mostly Goldstone excitons were formed. This enabled us
to select the required filling factor as well as to ensure quantum spin beats through quantization of the spin projection
($S_{z}$) in the magnetic field direction.

\vspace{-4mm}

\begin{figure}[!h]
\includegraphics[width=\linewidth]{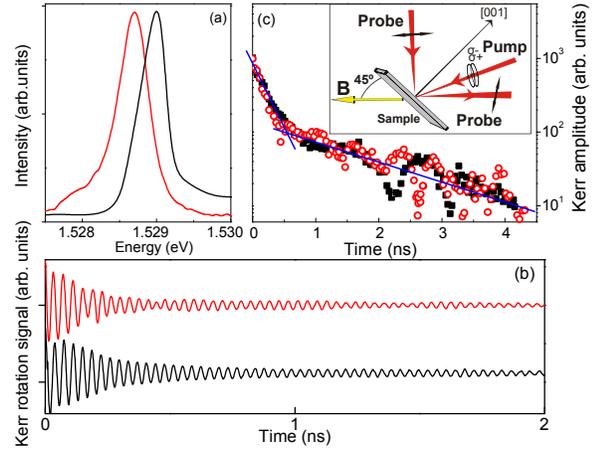}
\vspace{-8mm}\caption {(a) Photoluminescence spectrum corresponding to optical transition ($1/2;-3/2$) (red curve) and ($-1/2;3/2$) (black curve)
from the lowest Landau level of the conductance band to the lowest Landau level of the heavy holes in the valence band
at magnetic field 4.3~T ($\nu$  = 0.96). (b) Kerr signal involving optical transitions ($1/2;-3/2$) (red curve) and
($-1/2;3/2$) (black curve). (c) Time behavior of Kerr signal amplitude. The straight lines are obtained by means of double
exponential approximation of the experimental points.
The insert schematic illustrates a 45 degree tilted-field geometry for time- and spectrally-resolved spin Kerr-effect measurements.}\vspace{-3mm}
\end{figure}

The beating amplitude (the difference between the maximum and minimum of the time- and spectrally-resolved spin Kerr-effect signal) decays at two different times, short $T_{S1}^{e}$ (several hundreds of picoseconds) and long $T_{S2}^{e}$ (nanoseconds) (Fig.2). {In  addition, a beating signal is modulated by low-frequency oscillations. Those are observed in highest mobility samples only and disappear in samples with mobility lower than $3\!\times\!10^{6}\,$cm$^{2}$/Vs. We attribute these oscillations to influence of a plasma vibration of the whole electron system, the origin of which is yet unknown.}
The prime relaxation time $T_{S1}^{e}$ is independent of the filling factor, while
the dependence of $T_{S2}^{e}$ on $\nu$ becomes dramatic near the ferromagnet values $\nu\!=\!1,\,3$. Since the initial phase relaxation ($T_{S1}^{e}$) is not related to the filling factor, it is attributed to single electron spin relaxation. The electron system is likely to be overheated immediately after the pumping pulse, and relaxation time $T_{S1}^{e}$ emerges due to cooling \cite{Plochocka09}. This assumption is supported by the fact that increasing the pumping power enhances the fast relaxing part of the Kerr precession.
Below, we consider the long-time relaxation channel only, which is sensitive to the spin arrangement of the ground state (Fig.3).

\vspace{-2.mm}

\begin{figure}[!h]
\includegraphics[width=\linewidth]{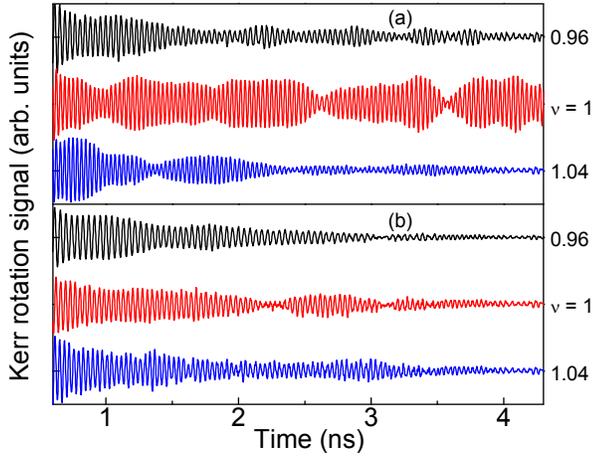}
\vspace{-6mm}
\caption {Long-lived part of Kerr signal registered in
spectral position corresponding to lower energy optical transition (a) and to higher energy optical transition (b) (see Fig.1)
as a function of the filling factor near $\nu\!=\!1$.}\vspace{-3mm}
\end{figure}

To illustrate how different energy states participate in the spin relaxation dynamics, we present the Kerr signal obtained from pumping in resonance with two optical transitions: (i) from the valence band to the electron Fermi edge, (Fig.~3, top panel) and (ii) from the valence band to the maximum density of the empty states ($\sim\!0.5\,$meV higher: Fig.~3, lower panel). The spin dynamics of the higher energy excitation was only slightly sensitive to the filling factor (Fig.${}\,$3b), while the spin dephasing time for the electrons at the Fermi edge was dramatically longer (Fig.${}\,$ 3a).
When $\nu$ is exactly equal to 1, the spin dephasing time can be estimated only
roughly. Since the periodicity of the laser pulses is $12.5\,$ns, time intervals
longer than $10\,$ns can hardly be measured with our technique.
Given the tiny deviation from the exact value $\nu\!=\!1$, the spin dephasing time
decreased by more than an order of magnitude.

For reference: the samples investigated in the previous Kerr experiments \cite{Fukuoka} could not really exhibit collective states as the mobility of the samples $\mu_{e} \simeq 1\!\times\!10^{6}\,$cm$^{2}$/Vs does not imply the presence of a long-range order QHF. In those works the change of spin relaxation time in transition from the QHF to skyrmion systems demonstrated a tiny change in spin relaxation time from 6 to 4 ns. It means that there is neither a QHF nor a skyrmion crystall/liquid. The small change in relaxation time exhibits the influence of the weak collective effects on the basically single-particle physics in an inhomogeneous system. The existing theories predict a much larger change in spin relaxation time when the electron system is driven from a QHF to a skyrmion texture. For instance, nuclear spin relaxation under similar conditions involves variations in relaxation time up to two orders of magnitude \cite{Tycko}, and the theoretically predicted time variations could reach three orders of magnitude \cite{Fertig96}. {(Both nuclear and electron spins relax through the same electron excitations. The absolute relaxation times for nuclei may exceed those for electrons by many orders of magnitude. Yet, due to the transition
from the ferromagnetic to the skyrmion system the relative changes should be similar for both systems.)} The strong variation of the relaxation time is the key proof of the quantum phase transition from a QHF to a skyrmion system.  We observe variations of relaxation time over a narrow range around filling  $\nu\!=\!1$ up to 20-fold which points to the presence of the phase transition in question. The striking difference between our and the previous experimental results is likely to be due to the fairly higher quality samples used in the present study as well as the application of spectrally resolved Kerr rotation, as the change of the spectral characteristics of the excitation beam leads to a change in spin relaxation time by an order of magnitude (Fig.~4). {And also a comment on the discussion presented in Ref. \cite{Fukuoka} which is based on theoretical work \cite{di04}: in Ref. \cite{di04} another initial state was studied and hypothetically a single mechanism was considered for both stochastization and relaxation processes corresponding to the simultaneous $S_z\!\to\!S_z\!+\!1$ and $S\!\to\!S\!-\!1$
transitions (see the introductory part above). This approach is irrelevant to any explanation of the observed Zeeman-frequency precession and leads to an estimate for the relaxation time definitely longer than $10\,$ns (really longer than $100\,$ns in an up-to-date quantum well). So, there is no agreement with time $\!\approx\!6\,$ns \cite{Fukuoka}. In our present study we assert that the stochastization ($S\!\to\!S\!-\!1$ under condition
$S_z\!=const$) and the relaxation ($S_z\!\!\to\!S_z\!+\!1$ under condition $S\!=\!S_z$) are determined by physically different mechanisms.}

The huge variability in the electron relaxation time around $\nu\!=\!1$ is consistent with the results of Ref. \cite{Tycko} regarding the influence of the spin rearrangement on nuclear spin relaxation. The reason is as follows: when the electron system undergoes the phase transition from the QHF to a less rigid spin state, several phase-destroying mechanisms for the coherent spin precession come into play due to low energy spin excitations \cite{ga08,dr10,Fertig96}.
\vspace{-0.mm}
\begin{figure}[!h]
\includegraphics[width=\linewidth]{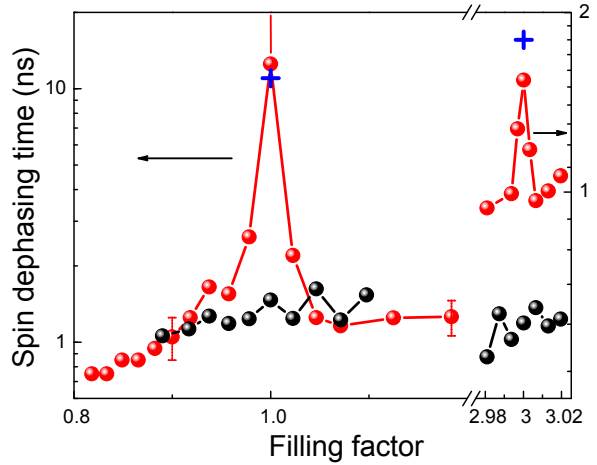}
\vspace{-7.mm}
\caption {Decay time of long-lived Kerr signal registered in
spectral position corresponding to lower energy optical transition (red) and to higher energy optical transition (black) (see Fig. 1)
at different filling factors near of $\nu\!=\!1$ (total magnetic field $B\!=\!4.50\,$T, sample A) and of $\nu\!=\!3$ ($B\!=\!4.65\,$T, sample B). Blue crosses correspond to
theoretical estimations (see the text). }\vspace{-6.5mm}
\end{figure}

{\em Discussion.} The Kerr signal, measured at a moment $t$, is proportional to the quantum-mechanical average of the non-conserved value $S_x\!+\!iS_y$. However
both states $|N,0\rangle$ or $|N,{\bf q}\rangle$ give
$\langle{\hat S}_x\!+\!i{\hat S}_y\rangle\!\equiv\!0$.
To study the Kerr precession, we first describe the time-dependent states.
The initial state arises as a result of a very fast
`vertical' recombination transition induced by light absorbtion. Due
to elementary single-photon annihilation, instead of an up-spin electron
$\uparrow=\!\left({1\atop 0}\right)$ a `tilted' electron {\footnotesize ${\nearrow}$}${}\!=\!\!\left({\cos{\!\frac{\beta}{2}} \atop  -\sin{\!\frac{\beta}{2}}}\right)$ emerges, where $\beta$ is one of the Euler angles (two others may be equated to zero without
loss of generality; $\beta$ is close to $45^{\rm o}$ in the experiment). If
one photon is absorbed in the state $|0,0\rangle$, the initial state represents a combination
of vectors $|${\footnotesize ${\nearrow}\,$}$\uparrow\uparrow\uparrow\!...\!\uparrow\rangle$, $|\!\uparrow${\footnotesize${\nearrow}\,$}$\uparrow\uparrow\!...\!\uparrow\rangle$, ..., and
$\!|\!\uparrow\uparrow\uparrow\!...\!${\footnotesize${\nearrow}$}$\rangle$. Simple physical
considerations based on the indistinguishability principle lead to the following description
of the initial QHF state $|{\rm i}\rangle\!=\!{\hat L}_\beta(0)|0,0\rangle$, where
${\hat L}_\beta(0)\!=\!\cos{\!\frac{\beta}{2}}{\hat I}\!-\sin{\!\frac{\beta}{2}}{\cal Q}^\dag_0$ (${\hat I}$ denotes the unit operator). Here the appearance of a zero-exciton operator is stipulated by the strict `verticality' of the transition process, which held definitely in the experiment since ${\cal L}k_{{\rm phot}\parallel}\!\ll\!1$ ($k_{{\rm phot}\parallel}$ is the parallel photon momentum component and ${\cal L}$ is a linear characteristic of 2D density spatial fluctuations). The initial state is certainly not an eigenstate. It does not correspond to definite $S_z$, but still corresponds to definite $S\!=\!{\cal N}_\phi/2$. Under the experimental conditions, $N\!\ll\!N_{\phi}$, the elementary dephasing process is a single exciton process.

For simplicity, we consider a domain of area $A$ smaller than ${\cal A}_{\rm sp}/N$,
where ${\cal A}_{\rm sp}$ is the area of the laser spot. Accordingly, the Landau level
degeneracy is defined as ${\cal N}_\phi\!=\!A/2\pi l_B^2$ though certainly
assumed to be large, ${\cal N}_\phi\!\gg\!1$. It is clear that
no more than a single photon is absorbed within the $A$ domain, therefore our task is
to study the temporal evolution of the initial state $|{\rm i}\rangle\!=\!\cos{\!\frac{\beta}{2}}|0,0\rangle\!-\!\sin{\!\frac{\beta}{2}}|1,0\rangle$. In the absence of any violation of the translation invariance, the Schr\"odinger equation
results in state $|t\rangle\!=\!{\hat L}_\beta(t)|0,0\rangle$ at moment $t$,
where ${\hat L}_\beta(t)\!=\!\cos{\!\frac{\beta}{2}}{\hat I}\!-\sin{\!\frac{\beta}{2}}e^{-i\epsilon_Zt}{\cal Q}^\dag_0$. The calculation of expectation $\langle t|{\hat S}_x\!+\!i{\hat S}_y|t\rangle\!=\!-\frac{1}{2}\sin\!\beta\sqrt{{\cal N}_\phi}e^{-i\epsilon_Zt}$ explains the Kerr signal oscillations with frequency
$\epsilon_Z/\hbar$, but does not explain the Kerr signal decay.

To study the decay, we have to consider the stochastization process, which is slow
compared to the precession. This is a conversion of component $e^{-i\epsilon_Zt}|1,0\rangle$ of state $|t\rangle$ to component $e^{-i\epsilon_Zt}|1,{\bf q}\rangle$ at $q\!\to\!0$. Indeed, when calculating the $S_x\!+\!iS_y$ quantum average, any state $|t,{\bf q}\rangle\!=\!\cos{\!\frac{\beta}{2}}|0,0\rangle\!-\!e^{-i{\cal E}_qt}\sin{\!\frac{\beta}{2}}|1,{\bf q}\rangle$ is substituted for $|t\rangle$, and we come to a zero result: $\langle {\bf q},t|{\hat S}_x\!+\!i{\hat S}_y|t,{\bf q}\rangle\!\equiv\!0$. (Here ${\cal E}_q\!=\!\epsilon_Z\!+q^2\!/\!2\!M_{\rm x}$ is the spin exciton energy at small dimensionless $q$.) Thus the time of the Kerr signal decay is equal to the transition time of zero exciton $|1,0\rangle$ conversion into nonzero one $|1,{\bf q}\rangle\!{}_{q\to 0}$ with the same energy, ${\cal E}_0\!=\!\epsilon_Z$.

The perturbation responsible for the $|1,0\rangle\to |1,{\bf q}\rangle$
conversion must be: (i) a spin non-conserving coupling changing the $S$, but not changing
the $S_z$ quantum numbers; and (ii) violating the translation invariance.
The most likely candidate is a term corresponding to the spatial fluctuations
of the $g$-factor in 2D electron gas, i.e., the Zeeman energy is actually $\epsilon_Z\!+\!g_1({\bf r})\mu_BB$, where $\langle g_1\rangle\!\equiv\!\int
g_1({\bf r})d{\bf r}/A=0$ \cite{g-factor}.
For estimation, let us assume that the $g$-disorder is Gaussian and it is governed by correlator $K({\bf r})\!=\!\int\!g_1({\bf r}_0)g_1\!({\bf r}_0\!+\!{\bf r})d{\bf r}_0/A$, parameterized by fluctuation amplitude $\Delta_g$ and correlation
length $\Lambda_g$, i.e. $K({\bf r})\!=\!\Delta_g^2e^{-r^2/\Lambda^2_g}$. After performing some
manipulations similar to those described in Ref. \cite{di04} (where an electrostatic random potential was considered a dissipative mechanism), one finds that the stochastization occurs exponentially at a rate equal to
\vspace{-1.5mm}
\begin{equation}
\vspace{-1.5mm}1/\tau\!=\!\pi M_{\rm x}(\mu_BB\Delta_g\Lambda_g)^2\!/2\hbar l_B^2.
\end{equation}
\noindent{\vspace{-.5mm}Here $M_{\rm x}$ is the physical quantity responsible for the many-particle Coulomb/exchange coupling (the stronger the coupling $\sim\!e^2/\kappa l_B$, the smaller the spin-exciton mass \cite{Bychkov81,di04}). Formally Eq. (1) expresses
the following result: at equal magnetic fields and disorder parameters the Goldstone mode in a `rigid' ferromagnet (with realized large Coulomb constant, for instance, via small  dielectric constant $\kappa$) is more stable than in a `softer' one (with a larger $\kappa$)}.

Numerical estimation of $\tau$ is fairly complicated due to the scant information on the $\Delta_g$ and $\Lambda_g$ values. We estimate $\Delta_g/g\!\sim\!0.02$ for the $\nu\!=\!1$ sample, $\Delta_g/g\!\sim\!0.05$ for the $\nu\!=\!3$ one, and $\Lambda_g\!\!\sim\!50$~nm in both cases. Then using our knowledge of the spin-exciton mass [18] we can calculate $\tau$. These theoretical estimations are marked with crosses in Fig.~4.

So, the Goldstone mode stochastization is a crucially many-particle process which is different from the single-particle view of spin relaxation. We report on time $\tau$ corresponding to ``transverse'' time $T_2$ for spin relaxation. ``Longitudinal'' relaxation time $T_1$ characterizing the relaxation processes $\delta S_z\!\to\!0$ in a QHF was measured directly \cite{Zhuravlev14}. In accordance with spin relaxation physics of classical magnets, we find that the inequality $T_2\!\ll\!T_1$ also holds for the QHF.

It should be emphasized that optical creation of the Goldstone mode can be realized, in principle, not only in a GaAs/AlGaAs quantum Hall system. In future, a study similar to the above could probably be realized, e.g., in a QHF based on graphene, HgTe and ZnO/MgZnO systems.  Generalizing from the data obtained: the studied stochastization process should have a common nature for any 2D ferromagnets formed by purely conduction-band electrons.

This research was partially supported by the Russian Foundation for Basic Research.

\end{document}